\documentstyle[12pt]{article}

 1
 1
\begin{document}
%%%%%%%%%%%%%%%%%%%%%%%%%%%%%%%%%%%%%%%%%%%%%%%%%%%%%%%%%%%
\begin{titlepage}
\noindent
\begin{center}
\begin{Large}
\begin{bf}
SUSY Magnetic Moments Sum Rules \\
and Supersymmetry Breaking\\
\end{bf}
\end{Large}

\vspace{1.5cm}
\begin{large}
A.Culatti\footnote{e-mail address: culatti@padova.infn.it or
culatti@mvxpd5.pd.infn.it

Phone number 0039 (49) 831739 Telefax number 0039 (49) 844245} \\[3mm]
Dipartimento di Fisica Galileo Galilei\\
Universit\`a di Padova\\
Via Marzolo 8, 35131 Padova, Italy\\[3mm]
INFN, sezione di Padova, Italy
\end{large}

\vspace{1.0cm}
(revised version)

%
% Abstract
%

\vspace{1.5cm}
{\bf Abstract}
\end{center}
\vspace{0.2cm}
\begin{quotation}
\noindent
It was recently shown that unbroken N=1 Susy relates, in a
model independent way, the magnetic transitions between states of different
spin
within a given charged massive supermultiplet.
We verify explicitly these sum rules for a vector multiplet in the case of
massless and massive fermions.
The purpose of this analysis is to provide the ground for the broken susy case.
We study the modifications of these results when an explicit
soft Susy breaking realized through a universal mass for all scalars
is present.
As a by-product we provide a computation of the
$g-2$ of the $W$ boson in the standard model which corrects previous
evaluations in the literature.

\end{quotation}
\vspace{2.0cm}
DFPD/94/TH/25 \hfill june 1994
\end{titlepage}

\section{Introduction}

It was recently observed that in a $N=1$ supersymmetric (Susy) invariant
theory, the anomalous magnetic transitions among members of a vector or
higher spin supermultiplet are related by model independent sum rules~
\cite{fer-por} . When we consider a vector supermultiplet these rules
are very simple.
Indeed, calling $h$ the $g-2$ value of the charged vector bosons $W^{\pm}$,
it turns out that the anomalous magnetic moments of the fermionic
partners of $W$ (the charginos) are equal to $2h$ and the anomalous
magnetic transition between $W$ and its scalar partner (the charged higgs
$H^{\pm}$) is again equal to $h$.

The relevant question that we wish to address in this paper is the impact of
the breaking of supersymmetry on these sum rules.
In other words, one can try to use the abovementioned anomalous magnetic
moments sum rules as an indicator of the amount of susy breaking which
is present.

In order to perform this analysis we first make an explicit derivation of the
sum rule in the minimal supersymmetric standard model (MSSM) in the situation
of unbroken Susy. This derivation accomplishes a twofold purpose:
first it prepares the ground for an analysis of the departure from the exact
sum rule when Susy is broken in different ways, and, then, it allows for a
quick reappraisal of the results concerning the anomalous magnetic moment
of the $W$ boson in the standard model (SM).
We will show that this reanalysis leads to a sizeable correction of the
results previously reported in the literature for this computation.

Unless otherwise specified, our notations and conventions
are as in~\cite{bil-gas} where an exactly
supersymmetric version of the Weinberg-Salaam $ SU(2)_L\times U(1)_Y $ SM is
illustrated. In this model the charged massive vector multiplet of weak
interactions contains besides $ W^{\pm} $ gauge bosons and the Higgs scalars
$ H^{\pm} $, two spin-1/2 dirac fermions $ \omega^{-}_{1} $ and
$ \omega^{+}_{2} $ given by the linear combination of winos and higgsinos

\begin{equation}
\omega_1 = \sqrt{2} P_L\chi_{12} - iP_R\lambda^-
\end{equation}

\vspace{5mm}
\begin{equation}
\omega_2 = \sqrt{2} P_L\chi_{21} - iP_R\lambda^+.
\end{equation}
\vspace{5mm}

Here the Majorana fermions $ \chi_{12} $, $ \chi_{21} $ and $ \lambda^{\pm} $
are the supersymmetric partners of $ H^- $, $ H^+ $ and $ W^{\pm} $.
If Susy is unbroken all these particles have a common mass $ m_W $.
Gauge invariance is broken by an Higgs sector composed by two Higgs doublets
with opposite ipercharge $\pm 1$, needed to give mass in a Susy invariant
mode to both up and down quarks, and by an Higgs singlet N whose
ipercharge is zero.
Once Goldstone bosons are absorbed as the longitudinal degree of freedom
of $ W^{\pm} $ and $ Z^0 $ gauge bosons, we remain with 7 physical Higgs
scalars: $ H^{\pm} $, $ H^0 $, $ h^0_i $ ($ i $ =1,2,3,4).
$ H^{\pm} $ and $ H^0 $ have the same mass as $ W^{\pm} $ and $ Z^0 $ while
$ h^0_1 $,...,$ h^0_4 $ have a common mass $ m_h $.
As far as the particle content is concerned, the proliferation of Higgs
bosons is the only difference from the mere supersymmetrization of SM.

In a renormalizable theory of spin-$\frac{1}{2}$ and spin-1 particle, the tree
level value of the gyromagnetic ratio is 2. This is strictly tied with tree
level unitarity \cite{wei,fer-tel}.
However quantum effects can spoil this property. In particular,
for vector multiplets, $h$ could be nonzero due to loop effects.

Susy implies a strict relation between the following couplings

\vspace{5mm}
\[ \frac{g}{m_W} \overline{\omega}_i \sigma_{\mu \nu} \omega_i F^{\mu \nu}
\; \; \; \; \; \; \; \; \; \; \; \; %alternativo a%
\; \; \; \; \; \; \; \; \; \; \; \;
gW^+_{\mu}W^-_{\nu}F^{\mu \nu}
\]

\[ \frac{g}{m_W} \epsilon^{\mu \nu \rho \sigma} (\partial_{\mu} H^+W^-_{\nu}-
\partial_{\mu} H^-W^+_{\nu})F_{\rho \sigma}. \]
\vspace{5mm}

The most general CP and $U(1)_{e.m.}$ invariant $WW\gamma $ vertex when all
particle are on mass shell is \cite{bar-gas,hag-pec}

\begin{equation}
M_{\mu \alpha \beta}=ie\{A[2p_{\mu}g_{\alpha \beta}+4(q_{\alpha}g_{\beta \mu}-
q_{\beta}g_{\mu \alpha})]+2\Delta K_{WW}(q_{\alpha}g_{\beta \mu}-q_{\beta}
g_{\mu \alpha})+4\frac{\Delta Q}{m^2_W} p_{\mu}q_{\alpha}q_{\beta}\}
\label{wwgamma}
\end{equation}
\vspace{5mm}

where $p-q$, $p+q$, $2q$ are the momenta of the incoming and outgoing
$W^+$ and of the incoming photon.
In the standard model and in its supersymmetric version
at tree level $A=1$, while for the anomalous magnetic dipole and electric
quadrupole moments we have
$ \Delta K_{WW}=\Delta Q=0$.

The charginos $\omega_1 $ and $\omega_2 $ whose electric charges are
$e_{\omega_{1}}=-e$ and
$e_{\omega_{2}}=+e$ $(e>0)$ can have an anomalous magnetic moment
(in spite of chiral fermion $g_{1/2}=2$ in exact Susy theories)
$a_{\omega_{i}}=
\frac{g_{\omega_{i}}-2}{2}$, given by the coefficient of

\vspace{5mm}
\begin{equation}
\frac{1}{2m_W} e_{\omega_i}\bar{{\bf \omega}_i} \sigma^{\mu \nu}
q_{\nu}{\bf \omega}_i \varepsilon_{\mu} . \label{aomegai}
\end{equation}
\vspace{5mm}

with $q$ and $\varepsilon_{\mu}$ the momentum and polarization vector of the
incoming photon.
The presence of this term is due to the fact we can embed the anomalous
magnetic
moment of gauge fermion in a supersymmetric invariant term \cite{rob,bil-gas}
while we can not do the same thing for chiral fermions \cite{fer-rem}.

Besides this we must consider the off-diagonal magnetic
transition $\Delta K_{WH}$ between the spin-1 and spin-0 states in the
vector multiplet. It is characterized, when all particles are
on mass shell, as the coefficient of

\vspace{5mm}
\begin{equation}
\frac{e}{m_W} \varepsilon^{\mu \nu \rho \sigma} p_{\rho} q_{\sigma}
\varepsilon_{\mu} \varepsilon_{\nu}'
\end{equation}
\vspace{5mm}

where $p, \varepsilon_{\nu}' , q,\varepsilon_{\mu} $ are the moment and
polarization vector of the incoming $ W^+ $ and $ \gamma $ respectively.
Supersymmetric sum rules %~\ref{rule}
foresee

\vspace{3mm}
\begin{equation}
\Delta K_{WW}=a_{\omega_1}=a_{\omega_2}=\Delta K_{WH}.
\end{equation}
\vspace{5mm}

This forecast should be valid separately for the one loop contribution
due to particles owing to every single mass supermultiplet unless we try
to separate the contributions due to the exchange of quarks and leptons.
This last statement is strictly tied with anomaly cancellation as
proved in~\cite{bil-gas} .

We analyse the Susy magnetic moments sum rules for massless and massive
ordinary fermions in sections 2 and 3 respectively and the Susy breaking
case in section four.
We report the whole set of function we found in this last case in Appendix.
A comment on the size of the anomalous magnetic moment of $W$ in the
SM is given in the end of section 4 when we compare the SM with the exact
and broken susy
cases.
\section{Susy sum rules: The massless fermions case}

Bardeen et al. calculated $\Delta K_{WW}$ in the standard model
\cite{bar-gas,mik-sam} and Bilchak et al. showed
\cite{bil-gas} in the case of massless ordinary fermions
that $\Delta K_{WW}$ and $a_{\omega_i}$ have equal values. They verified this
is true also
if we consider separately the supermultiplets of photon, lepton, quarks, Z
boson and Higgs and found

\begin{eqnarray}
\Delta K^{\gamma}_{WW} = & a_{\omega_i} = & \frac{\alpha_{e.m.}}{\pi} \\
\Delta K^{l}_{WW} = & a^{l}_{\omega_1}+ &\frac{g^2}{32\pi^2} =
a^{l}_{\omega_2}-\frac{g^2}{32\pi^2} = 0  \label{dkl} \\
\Delta K^q_{WW} = & a^q_{\omega_1}- & \frac{g^2}{32\pi^2} =
a^q_{\omega_2}+\frac{g^2}{32\pi^2} = 0  \label{dkq} \\
\Delta K^Z_{WW} = & a^Z_{\omega_i} = & \frac{g^2}{16\pi^2} [\frac{4}{\rho}
-\frac{1}{2} +\int^{1}_{0} dx \frac{x^3-3x^2+4x-4}{x^2+\rho (1-x)}]\\
\Delta K^H_{WW} = & a^H_{\omega_i} = & \frac{g^2}{16\pi^2} [\frac{1}{2} +
\int^{1}_{0} dx \frac{x^2(1-x)}{x^2+\mu (1-x)} ]
\end{eqnarray}
\vspace{5mm}

Here $g$ is the weak coupling constant while
$\rho =(\frac{m_Z}{m_W} )^2$ and $\mu =(\frac{m_h}{m_W} )^2$.
We focalized our attention mainly on the contribution due to quarks, leptons
and their superpartners.

The diagrams contributing to $\Delta K_{WW}$ for one generation of quarks
and leptons are three, two with the up and down type quarks
running in the loop and the photon leg attached to one of these, and only
one for leptons because the neutrino is chargeless.
Three other diagrams have the same structure with the fermions
substituted by squarks and sleptons.
These graphs are illustrated in Fig.1-2.

As far as the contributions of the supermultiplets are considered, our
results agree with the Bilchak et al. ones \cite{bil-gas}.
However we find some difference as far as the separate quark and squark
contribution to $ \Delta K_{WW} $ are concerned.

We find for one generation of massless fermions

\begin{eqnarray}
\Delta K_{WW}(q)=\frac{-3g^2}{96\pi^2} \; \; \; \; \; \; \; \; \Delta K_{WW}
(\tilde{q} )=\frac{3g^2}{96\pi^2} \\
\Delta K_{WW}(l)=\frac{-g^2}{96\pi^2} \; \; \; \; \; \; \; \; \Delta K_{WW}
(\tilde{l} )=\frac{g^2}{96\pi^2}.
\end{eqnarray}
\vspace{5mm}

Indeed, if we call $N_c$ the number of
colours and $q_{u/d}$ the electromagnetic charges of the up and down
quarks in units of $e$ ($q_d=-\frac{1}{3}$, $q_u=\frac{2}{3}$),
we have\footnote{I wish to thank Prof. A. Van Proeyen for useful discussion
on this point.}

\vspace{3mm}
\begin{equation}
\Delta K_{WW}(q)=\frac{g^2}{96\pi^2} N_c(q_d-q_u) \label{disputa}
\end{equation}
\vspace{5mm}

This is a rather delicate point and therefore needs a careful analysis
to explain the relative minus sign between charges in equation~\ref{disputa}.
As we can see in Fig.1, the graphs with the photon attached to the up and down
quarks can be obtained one from the other with the substitutions
$ u\leftrightarrow d, \alpha \leftrightarrow \beta $ and
$ (p-q)_{in} \leftrightarrow (p+q)_{out} $. Last substitution is equivalent
to $ p \leftrightarrow -p $.
{}From this observations and from equation~\ref{wwgamma}
we deduce that if $ M^{\mu \alpha \beta}_{a}(u,d) $ is the vertex contribution
of graph a) with the $u,d$-dependence enclosed in the coefficients
$A(u,d), \Delta K_{WW}(u,d), \Delta Q(u,d)$, the vertex contribution of
graph b) is $ M^{\mu \alpha \beta}_{b}(u,d) =-M^{\mu \alpha \beta}_{a}(d,u) $.
This brings as a consequence that if the contributions of graph a) to
$ \Delta K_{WW} $ is $ \frac{g^2}{96\pi^2} N_cq_d $ (equal to the electronic
contribution), the sum with graph b) gives rise to equation~\ref{disputa}.

Analogously the same graphs produce the following contribution to the
electric quadrupole moment

\begin{equation}
\Delta Q(q)=\frac{-g^2}{72\pi^2} N_c(q_d-q_u). \label{minor}
\end{equation}
\vspace{5mm}

This reasoning can be applied to the charge renormalization coefficient $A$
and is valid for the sum of the squark loop graphs of Fig.2, too.
For these scalar graphs we obtain results opposite to~\ref{disputa} ,
\ref{minor} .

Nevertheless if we consider the anomalous term $ N^{\mu \alpha \beta} =
B\varepsilon^{\mu \alpha \beta \nu} p_{\nu} $ that can be generated by
fermion loops, it receives from graphs a) and b) of Fig.1
contributions $B_b(u,d) = B_a(d,u)$
without any change of sign because $\varepsilon^{\mu \alpha \beta \nu}
p_{\nu} $ is invariant under the previous set of substitutions.
Therefore, considering one complete fermion generation we have

\vspace{3mm}
\begin{equation}
B(q,l) = [N_c(q_d+q_u)+q_e]\cdot cost =0,
\end{equation}
\vspace{5mm}

so there is no problem for anomaly cancellation.

This point has been previously overlooked and, hence, our computation leads
to a different prediction for $\Delta K_{WW}$ in the standard model
even if it does not affect $\Delta K_{WW}$ in the massless fermions MSSM,
because quark and squark contributions cancel each other anyway.
Indeed, taking into
account the presence in SM of the three fermionic generations,
the final quarks and leptons exchange contribution is a sizeable one.

If we consider massless ordinary fermions, the chargino $\omega_1$
is only coupled to \cite{hab-kan} $d\tilde{u}_L$ and $e\tilde{\nu}_L$
and not to $u\tilde{d}_L$ and $\nu \tilde{e}_L$, with the opposite
assignation for $\omega_2 $.
These couplings give rise to the anomalous magnetic moment one loop corrections
illustrated in Fig.3-4. Here we list every single contribution
putting in evidence the particles whose propagators we met in the
loops.

\begin{eqnarray*}
a_{\omega_{1}}(d\tilde{u} \tilde{u} )   =  \frac{2g^2}{32\pi^2}
\; \; \; \; \; \; \; \;
a_{\omega_{2}}(u\tilde{d} \tilde{d} )  =  \frac{g^2}{32\pi^2} \\
a_{\omega_{1}}(dd\tilde{u} )   =  \frac{-g^2}{32\pi^2}
\; \; \; \; \; \; \; \;
a_{\omega_{2}}(uu \tilde{d} )  =  \frac{-2g^2}{32\pi^2} \\
a_{\omega_{1}}(ee\tilde{\nu} )  =  \frac{-g^2}{32\pi^2}
\; \; \; \; \; \; \; \;
a_{\omega_{2}}(\nu \tilde{e} \tilde{e} )  =  \frac{g^2}{32\pi^2}.
\end{eqnarray*}
\vspace{5mm}

Our results agree with \cite{bil-gas} .

In the massless fermions case we are considering, the vertex
$WH\gamma$ is one loop affected only by sleptons and squarks (coupling
$\tilde{u}_L \tilde{d}_L H =
\frac{-igm_W}{\sqrt{2}}$ and analogues)
since leptons and quarks have vanishing Yukawa couplings \cite{hab-kan}.
Besides this a scalar loop cannot provide terms like $\varepsilon^{\mu \nu
\rho \sigma} p_{\rho} q_{\sigma} \varepsilon_{\mu} \varepsilon_{\nu}' $, so
we have

\vspace{3mm}
\begin{equation}
\Delta K_{WH}^{q,\tilde{q} ,l,\tilde{l}}=0. \label{dkh}
\end{equation}
\vspace{5mm}

This completes  the verification of the sum rules for the
quark and lepton multiplets.

We list here the $\Delta K_{WH}$ contributions due to the $\gamma, Z, H$
supermultiplets illustrated in Fig.6, so, taking into account eqations 7, 10,
11, the verification is complete for the other multiplets, too.

\vspace{3mm}
\begin{eqnarray}
\Delta K_{WH}(\tilde{\gamma} \omega_i\omega_i) = &
\frac{1}{2}\Delta K^{\gamma}_{WH} = & \frac{\alpha_{e.m.}}{2\pi} \\
\Delta K_{WH}(\zeta \omega_i\omega_i) = &
\frac{1}{2}\Delta K^Z_{WH} = & \frac{g^2}{32\pi^2} [\frac{4}{\rho}
-\frac{1}{2} +\int^{1}_{0} dx \frac{x^3-3x^2+4x-4}{x^2+\rho (1-x)}]\\
\Delta K_{WH}(\tilde{h} \omega_i\omega_i) = &
\frac{1}{2}\Delta K^H_{WH} = & \frac{g^2}{32\pi^2} [\frac{1}{2} +
\int^{1}_{0} dx \frac{x^2(1-x)}{x^2+\mu (1-x)} ]
\end{eqnarray}
\vspace{5mm}

Here $\tilde{\gamma}, \zeta $ and $\tilde{h}$ are the fermionic partner
of $\gamma, Z$ and $h_i$ respectively.

\section{Susy sum rules: Massive fermions}

We can rely upon the results of the previous section as far as
the first two fermions generations are concerned,
because the fermions masses involved can be neglected compared
to $m_W$.
Nevertheless in a realistic model we must consider the fact that the
top quark must be heavier than $W$, too.

We shall consider two cases: a) $m_W$ negligible with respect to $m_t$ and
fixed
ratio $(\frac{m_b}{m_t} )^2=r$;
b) fixed ratio $(\frac{m_W}{m_t} )^2=\alpha$
with $m_b=0$.
Given the established hierarchy $m_b<<m_W<m_t$ (with $m_t$ of the order of
twice $m_W$  according to the precision tests of the standard model physics at
LEP (see for example \cite{lep-col}) and the first indication for direct top
evidence at Tevatron \cite{cdf-col} ), the approximation of case b) in which
terms of $O(\frac{m_b}{m_t} )$ are neglected is certainly more realistic.
In both cases we'll consider massless leptons.

Considering massive ordinary fermions brings as a consequence a multiplication
of the couplings appearing in the lagrangian. In particular we have
new vertices like $\bar{t} bH$ and the fermion number violating
$\omega_{1}^{c} \bar{t} \tilde{b}_R$ and
$\omega_{2}^{c} \bar{b} \tilde{t}_R$ \cite{gun-hab}.
These new couplings generate the diagrams illustrated in Fig.5-7-8 which
contribute to $\Delta K_{WH}$ and $a_{\omega_i}$.

Case a) was just analyzed in \cite{fer-mas} and we report here
only the final result valid for the contributions of the quarks and
leptons multiplets

\begin{equation}
\Delta K_{WW}^{ql}=a_{\omega_{1}}^{ql}=a_{\omega_{2}}^{ql}=\Delta K_{WH}^{ql}=
\frac{-g^2}{32\pi^2} F(r) \label{dkql}
\end{equation}
with
\begin{equation}
F(r)=
\frac{1}{(1-r)^3} [r^3+11r^2-13r+1-4r(1+2r)\ln{r} ].
\end{equation}

Here the result for the third generation coincides with the three
generations one.
This time we have an universal non null function $F(r)$ with which we can
express the whole set of anomalous magnetic moments. In the interesting
cases $m_b=0$, $m_t=0$, $m_b=m_t$ we have respectively $F(0)=1$, $F(\infty )=
-1$ and $F(1)=1$.

For case b) we report the whole set of functions we obtain.
In particular we find it relevant that our results restricted to the SM case
exhibit some difference with the values previously reported in the literature.
If we put $a=(\frac{m_t}{m_W} )^2$ and $b=(\frac{m_b}{m_W} )^2$ we obtain

\begin{eqnarray}
\Delta K_{WW}(bbt) & = & \frac{g^2N_cq_b}{32\pi^2}
\int_0^1 dx\frac{x^4+x^3(a-b-1)+x^2
(2b-a)}{bx+a(1-x)-x(1-x)}  \label{a} \\
\Delta K_{WW}(\tilde{b} \tilde{b} \tilde{t} ) & = & \frac{-g^2N_cq_b}{16\pi^2}
\int_0^1 dx\frac{(x^3-x^2)(b-a-1+2x)}{bx+a(1-x)-x(1-x)}
\end{eqnarray}
\begin{eqnarray}
a_{\omega_{1}}(b\tilde{t} \tilde{t}) & = & \frac{g^2N_cq_t}{16\pi^2}
\int_0^1 dx \frac{x(x-1)[b(x-2)+x]}{ax+b(1-x)-x(1-x)} \\
a_{\omega_{1}}(bb \tilde{t}) & = & \frac{g^2N_cq_b}{16\pi^2}
\int_0^1 dx \frac{x^2[b(x+1)+x-1]}{bx+a(1-x)-x(1-x)} \\
a_{\omega_{1}}(t\tilde{b} \tilde{b}) & = & \frac{g^2N_cq_b}{16\pi^2}
\int_0^1 dx \frac{x^2b(1-x)}{bx+a(1-x)-x(1-x)} \\
a_{\omega_{1}}(tt\tilde{b}) & = & \frac{g^2N_cq_t}{16\pi^2}
\int_0^1 dx \frac{x^2b(1-x)}{ax+b(1-x)-x(1-x)}
\end{eqnarray}

\vspace{2mm}
\begin{equation}
\Delta K_{WH}(tbb)=\frac{-g^2N_cq_b}{16\pi^{2}} \int^{1}_{0} dx
\frac{x[a(1-x)-bx]}{bx+a(1-x)-x(1-x)}. \label{f}
\end{equation}
\vspace{3mm}

We obtain
the corresponding contributions for $\Delta K_{WW}(btt)$, $\Delta K_{WW}
(\tilde{b} \tilde{t} \tilde{t} )$, $a_{\omega_2} (t\tilde{b} \tilde{b} )$,
$a_{\omega_2} (tt\tilde{b} )$,
$a_{\omega_2} (b\tilde{t} \tilde{t} )$,
$a_{\omega_2} (bb\tilde{t} )$,
$\Delta K_{WH} (btt)$,
with the substitutions $q_b\leftrightarrow -q_t$,
$m_b\leftrightarrow m_t$.

The relative minus sign between the charges $q_b$ and $q_t$ has already been
explained in the case of $\Delta K_{WW}$. As far as $a_{\omega_i}$ is
concerned it is simply due to $a_{\omega_i}$ definition~(~\ref{aomegai}~)
which contains $e_{\omega_i}$.
Writing explicitly the matrix elements for the two
$\Delta K_{WH}$ quark contributions
depicted in Fig. 5
and using usual trace properties, the same
substitution statement is easily verified.

Summing up the whole set of results~\ref{a}, ..., \ref{f}
and taking into account a massless
lepton generation we obtain

\vspace{3mm}
\begin{equation}
\Delta K_{WW}^{q\tilde{q} l\tilde{l}} = a_{\omega_1}^{q\tilde{q} l\tilde{l}} =
 a_{\omega_2}^{q\tilde{q} l\tilde{l}} = \Delta K_{WH}^{q\tilde{q} l\tilde{l}} =
\frac{g^2}{16\pi^2} \int_{0}^{1} dx \frac{(1-3x)(ax-b(1-x))}{ax+b(1-x)-x(1-x)}
\end{equation}
\vspace{2mm}

Inserting $b=0$
and expressing everything as a function of $\alpha =(\frac{m_W}
{m_t} )^2$ we find

\vspace{3mm}
\begin{equation}
\Delta K_{WW}^{q\tilde{q} l\tilde{l}} = a_{\omega_1}^{q\tilde{q} l\tilde{l}} =
 a_{\omega_2}^{q\tilde{q} l\tilde{l}} = \Delta K_{WH}^{q\tilde{q} l\tilde{l}} =
\frac{-g^2}{32\pi^2} G(\alpha )
\end{equation}
\vspace{2mm}

with

\vspace{1mm}
\begin{equation}
G(\alpha )=\frac{2}{\alpha^2} [3\alpha +(3-2\alpha )\ln (1-\alpha )].
\end{equation}
\vspace{3mm}

This function has
$ \lim_{\alpha \rightarrow 0} G(\alpha )=1$ so reproducing the case of
negligible $m_W$. Again the anomalous magnetic moments supersymmetric
sum rules are exactly verified.

The complete expression obtained using the realistic value $m_b=5GeV$ give
results which differ from those obtained with $G(\alpha )$ by O(10\%) if
$m_t\simeq 100GeV$ and only by O(1\%) if $m_t>160GeV$.

The function $G(\alpha )$ exhibit a divergence for $m_t=m_W$ because in this
and in the more general case $m_b+m_t=m_W$ the diagrams we studied have a
singularity in the physical region due to the presence of the threshold
for the $W\rightarrow tb$ decay.
Our calculation should be reliable provided $m_t$ differs from $m_W$ more than
the $W$ decay width.

If we limit ourselves to the SM contribution to $\Delta K_{WW}^{ql}$
due to the third generation, keeping
$m_t$ as the only non null fermion mass we have

\vspace{3mm}
\begin{equation}
\Delta K_{WW}^{ql}(SM)=\frac{-g^2}{96\pi^2} \frac{1}{\alpha^3}  [4\alpha^3-
3\alpha^2 +18\alpha +6(3-2\alpha )\ln (1-\alpha )]. \label{ksm}
\end{equation}
\vspace{3mm}

This result differs from those found with the same assumptions
in \cite{cou-ng}
and \cite{sam-sam}. Such difference can be traced back
to the erroneous summation of the two quarks diagrams analogously to
what happens in the massless case \cite{bar-gas,bil-gas} we clarified
before.

We note that equation~\ref{ksm} reduce to $\Delta K_{WW}^{ql}(SM)=
\frac{-g^2}{24\pi}$ if $\alpha$
goes to $0$ or $\infty $ and this is the same result we obtained in
the massless fermion case .
As a further ("a posteriori") check of validity of~\ref{ksm}
we observe that summing up
with the supersymmetric contributions we find the same universal function
$G(\alpha )$ that arises independently in the computation
of $a_{\omega_1}$, $a_{\omega_2}$
and $\Delta K_{WH}$.

\section{Soft breaking with scalar masses}

As a final step we calculated  the total contribution to the four quantities
we considered in the MSSM with Susy broken explicitly but softly by an
universal mass $\tilde{m}$ for every scalar particle we have in the theory,
to make a comparison with the unbroken case and search for a
possible new rule relating magnetic moments in broken Susy multiplets.

This choice of Susy breaking is not only the simplest but can also be seen
as one of the possible low energy remnant of string theory. Particularly
in a large class of string scenarios (symmetric orbifolds)
scalar masses should be largely bigger than gaugino masses
and in spite of a general lack of universality they should be
nearly universal because of the weak dependence
from their corresponding modular weights \cite{iba-lus}.

The tree level $\tilde{m} $ introduction effect is only a shift in scalar mass
eigenstates which does not affect directly any other coupling including
$\omega_i q_j\tilde{q}_k$, $q_iq_jH$ and $\tilde{q}_i \tilde{q}_j H$ which
depend only on fermion masses generated through their Yukawa couplings.

The whole set of results we obtained is reported in Appendix where the
dependence from the new parameter $\delta = (\frac{m_W}{\tilde{m}})^2 $
is put in evidence.
These results are not very enlightening and so
we examined the quark and lepton multiplet contribution
for 1 fermion generation in three particular
cases: i) massless fermions; ii) only one massive fermion (top);
iii) one heavy completely isomassive fermions generation.

In case i) the values reported in equations~\ref{dkl} , \ref{dkq}
and~\ref{dkh} for the unbroken Susy case get modified when $\tilde{m}>>m_W$:

\vspace{3mm}
\begin{equation}
\Delta K_{WW}^{q\tilde{q} l\tilde{l} } = \frac{-g^2}{24\pi^2}
\end{equation}
\begin{equation}
a_{\omega_1}^{q\tilde{q} l\tilde{l} }=a_{\omega_2}^{q\tilde{q} l\tilde{l} }
=\frac{g^2}{48\pi^2}
\end{equation}
\begin{equation}
\Delta K_{WH}^{q\tilde{q} l\tilde{l} }=0.
\end{equation}
\vspace{3mm}

In case ii) the corresponding starting value is $\frac{-g^2}{32\pi^2}$
(see equation~\ref{dkql} in the $r\rightarrow 0$ limit),
while for $\tilde{m}>>m_{t/W} $ the results are

\vspace{3mm}
\begin{equation}
\Delta K_{WW}^{q\tilde{q} l\tilde{l} } = \frac{-g^2}{24\pi^2}
\end{equation}
\begin{equation}
a_{\omega_1}^{q\tilde{q} l\tilde{l} } =
a_{\omega_2}^{q\tilde{q} l\tilde{l} }= \frac{g^2}{48\pi^2}
\end{equation}
\begin{equation}
\Delta K_{WH}^{q\tilde{q} l\tilde{l} } = \frac{-g^2}{32\pi^2}.
\end{equation}
\vspace{3mm}

If we change the mass hierarchy setting $m_t>>\tilde{m} >>m_W$ they become

\vspace{3mm}
\begin{equation}
\Delta K_{WW}^{q\tilde{q} l\tilde{l} } =a_{\omega_2}^{q\tilde{q} l\tilde{l} }=
\frac{-g^2}{24\pi^2}
\end{equation}
\begin{equation}
a_{\omega_1}^{q\tilde{q} l\tilde{l} } = \frac{-g^2}{96\pi^2}
\end{equation}
\begin{equation}
\Delta K_{WH}^{q\tilde{q} l\tilde{l} } = \frac{-g^2}{32\pi^2}.
\end{equation}
\vspace{3mm}

In the final case of a completely isomassive generation
we start from $\frac{-g^2}
{24\pi^2}$ and in the case of $\tilde{m}>>m $ with $m$ the common fermion
mass we have

\vspace{3mm}
\begin{equation}
\Delta K_{WW}^{q\tilde{q} l\tilde{l} } =\Delta K_{WH}^{q\tilde{q} l\tilde{l} }
= \frac{-g^2}{24\pi^2}
\end{equation}
\begin{equation}
a_{\omega_1}^{q\tilde{q} l\tilde{l} } =
a_{\omega_2}^{q\tilde{q} l\tilde{l} } = \frac{g^2}{48\pi^2}.
\end{equation}
\vspace{3mm}

Instead of these if we set $m>>\tilde{m} >>m_W$ we obtain

\vspace{3mm}
\begin{equation}
\Delta K_{WW}^{q\tilde{q} l\tilde{l} } =\Delta K_{WH}^{q\tilde{q} l\tilde{l} }
= a_{\omega_1}^{q\tilde{q} l\tilde{l} } =
a_{\omega_2}^{q\tilde{q} l\tilde{l} } = \frac{-g^2}{24\pi^2}.
\end{equation}
\vspace{3mm}

As we can see from these results we haven't a common clear sign relating
Susy breaking with the new sum rules we could write in every single case.
The problem is even more complicated if we consider non negligible $m_t$
or $m_{W}$ compared to $\tilde{m}$, in which case the four magnetic
moments we considered have four different and apparently uncorrelated
values.
Indeed this happens with an explicit Susy breaking but we have a first
indication things can hardly go better in the case of spontaneous breaking
because quark and lepton supermultiplet contributions should always be
present and they are rather independent from the contributions of the
other multiplets (the fermion mass hierarchy is not fixed by Susy).

We conclude our work with some tables giving the total results for
$\Delta K_{WW}$ considering the whole set of diagrams contributing to it,
$\gamma$, $Z$ and $H$ multiplets included,
in the case of the SM (Table 1), of its minimal exactly supersymmetric
version (Table 2) and in the case
of Susy broken by the universal scalar mass $\tilde{m}$.

In the case of SM and its minimal exactly supersymmetric version
we give the results in function of
$\mu=(\frac{m_h}{m_W} )^2 $ and $\alpha = (\frac{m_W}{m_t} )^2$.
In the case of broken Susy they are expressed in function of $\mu$ and
$\delta = (\frac{m_W}{\tilde{m}} )^2$ at fixed values of $\alpha$ belonging
to the favourite range 100-200GeV for $m_t$.

Throughout these computations we used $\sin^2{\theta_W }=.2325$
and $\rho =(\frac{m_Z}{m_W} )^2 =1.29$.

The standard model results differs significantly from the published
one because of the non cancellation of the fermion contributions
(three generations).
The most striking result is that $\Delta K_{WW}$ is negative, unless
a light Higgs is present, and rather small.
This result contrasts with what was
previously found in the literature.
As we can see in Table 1, $\Delta K_{WW}$ is a decreasing function both
of $m_t$ and $m_h$.
If we assume the CDF result~\cite{cdf-col} $m_t=174\pm 17$ GeV and use the
lower
bound $m_t=157$ GeV we find the largest $m_h$ that allows a positive
$\Delta K_{WW}$. We find

\vspace{3mm}
\begin{equation}
\Delta K_{WW}>0 \Longrightarrow    m_h\leq 81 GeV
\end{equation}
\vspace{3mm}

It must be remembered the LEP1 limit~\cite{sch} $m_h>63.8$ GeV that can
be raised to $80-90$ GeV at LEP2.

Unfortunately the corrections to $g_W=2$ remains always of the order
$10^{-3}$, both in the standard and in the supersymmetric case, making
impossible an experimental verification as
today we know only $g_W$ is $O(1)$ and in the near future we will
probably know it at best with a precision ten times better \cite{sam-li}.

The results reported in Table 3 should be taken as an indication of the
departure from the unbroken Susy case in the simplest possible approach
where the entire Susy breaking is accounted for by a universal scalar
mass. Clearly a more detailed analysis should take into account the influence
of a specific breaking of supergravity at a large scale down to low energy
through the evolution of the full Susy parameter spectrum.

A final comment on the size of the contributions in the tables is in order.
It's well known that the supersymmetric corrections to the $g-2$ of the muon
in MSSM
is particularly small \cite[and references therein]{hab-kan} and, indeed,
it does not lead to significant constraints on the sparticle masses given the
present experimental lower bounds on them. On the contrary the supersymmetric
contributions to the $g-2$ of the $W$ boson is of the same order of magnitude
as the SM corrections. This is due to the fact that the ratio of the $W$
and Susy masses is much larger than the analogous ratio of the muon to the Susy
masses.

As we can see from table 2, exact Susy predict a rather different range
of values compared with the SM one, resulting
in a probable difference of sign between them. However
the explicit supersymmetry breaking with $\tilde{m} >m_W$ (phenomenology
states Susy
breaking cannot be to much little) modifies
in a sizeable way the squarks and sleptons contribution and the $H^+H^0$,
$W^+h^0_1$, $H^+h^0_2$ ones depicted in Fig.9
\cite{bil-gas}. This leads to foresee results
nearer to the SM and with the same sign.

\vspace{1.2cm}
{\bf Acknowledgment}

\vspace{0.5cm}
I wish to thank Antonio Masiero for his guide and patient support,
Antoine Van Proeyen for his kind interest and Ferruccio Feruglio
for useful discussions.

\newpage
\begin{Large}
\begin{bf}
Appendix \hfill
\end{bf}
\end{Large}
\vspace{10mm}

We report the whole set of scalar
contributions to $\Delta K_{WW}$ modified by
the introduction of the universal Susy breaking parameter $\tilde{m}$
as a function of
$a_i=(\frac{m_{u_i}}{m_W} )^2$,
$b_i=(\frac{m_{d_i}}{m_W} )^2$
and $\delta =(\frac{m_W}{\tilde{m}} )^2$ with $u_i=u,c,t$ and $d_i=d,s,b$

\vspace{3mm}
\begin{eqnarray*}
\Delta K_{WW}(\tilde{d}_i \tilde{d}_i \tilde{u}_i ) & = &
\frac{g^2N_cq_{d_i}}{16\pi^2} \int_{0}^{1} dx \frac{\delta (x^2-x^3)
(b_i-a_i-1+2x)}{b_ix+a_i(1-x)+\frac{1}{\delta} -x(1-x)} \\
\Delta K_{WW}(\tilde{d}_i \tilde{u}_i \tilde{u}_i ) & = &
\frac{-g^2N_cq_{u_i}}{16\pi^2} \int_{0}^{1} dx \frac{\delta (x^2-x^3)
(a_i-b_i-1+2x)}{a_ix+b_i(i-x)+\frac{1}{\delta} -x(1-x)}.
\end{eqnarray*}
\vspace{5mm}

Here and in the following,
leptonic contributions can be easily obtained from those which contains
$q_{d_i}$ with the substitutions $N_cq_{d_i} \rightarrow q_{l_i}, a_i
\rightarrow a_i'$ and $b_i \rightarrow b_i'$ with
$a_i' = (\frac{m_{\nu_i}}{m_W} )^2$ and $b_i' = (\frac{m_{l_i}}{m_W} )^2$.

Considering quark $t$ as the only
massive fermion and setting $\alpha =(\frac{m_W}{m_t} )^2$,
the scalar quarks and leptons give

\vspace{3mm}
\begin{eqnarray*}
\Delta K_{WW}(\tilde{d} \tilde{d} \tilde{u} ) & = &
K_{WW}(\tilde{s} \tilde{s} \tilde{c} ) =
K_{WW}(\tilde{e} \tilde{e} \tilde{\nu} ) =
K_{WW}(\tilde{\mu} \tilde{\mu} \tilde{\nu} ) =
K_{WW}(\tilde{\tau} \tilde{\tau} \tilde{\nu} )  \\
& = & \frac{-g^2}{16\pi^2} \int_{0}^{1} dx \frac{\delta (x^2-x^3)(2x-1)}
{1-\delta x(1-x)} \\
\Delta K_{WW}(\tilde{d} \tilde{u} \tilde{u} ) & = &
K_{WW}(\tilde{s} \tilde{c} \tilde{c} ) =
2K_{WW}(\tilde{d} \tilde{d} \tilde{u} )  \\
%& = & \frac{-2g^2}{16\pi^2} \int_{0}^{1} dx \frac{\delta (x^2-x^3)(2x-1)}
%{1-\delta x(1-x)}  \\
\Delta K_{WW}(\tilde{b} \tilde{b} \tilde{t} ) & = &
\frac{-g^2}{16\pi^2} \int_{0}^{1} dx \frac{\delta (x^2-x^3)(\alpha (2x-1)-1)}
{\alpha +\delta (1-x)-\alpha \delta x(1-x)}  \\
\Delta K_{WW}(\tilde{b} \tilde{t} \tilde{t} ) & = &
\frac{-2g^2}{16\pi^2} \int_{0}^{1} dx \frac{\delta (x^2-x^3)(\alpha (2x-1)+1)}
{\alpha +\delta x-\alpha \delta x(1-x)}.
\end{eqnarray*}
\vspace{5mm}

The ordinary Higgs contributions get modified by $\tilde{m} $ introduction,
too.
The functions reported in \cite{bil-gas} become

\vspace{3mm}
\begin{eqnarray*}
\Delta K_{WW}(H^+H^+H^0) & = &
\frac{g^2}{16\pi^2} [\frac{1}{6} +\frac{1}{2} \int_{0}^{1} dx \frac{x^2(x^2
-2x-\frac{1}{\delta})}
{x^2+\rho (1-x)+\frac{1}{\delta} } ] \\
\Delta K_{WW}(W^+W^+h^0_1) & = &
\frac{g^2}{16\pi^2} [\frac{1}{6} +\frac{1}{2} \int_{0}^{1} dx \frac{x^2(x^2
-2x+4)}
{x^2+(\mu +\frac{1}{\delta } ) (1-x)} ] \\
\Delta K_{WW}(H^+H^+h^0_2) & = &
\frac{g^2}{16\pi^2} [\frac{1}{6} +\frac{1}{2} \int_{0}^{1} dx \frac{x^2(x^2
-2x-\frac{1}{\delta })}
{x^2+\mu (1-x)+\frac{1}{\delta }} ],
\end{eqnarray*}
\vspace{5mm}

where we used $\rho = (\frac{m_Z}{m_W} )^2 $ and $\mu = (\frac{m_h}{m_W} )^2 $.
We recover exactly supersymmetric functions simply by letting $\delta$ goes
to infinity.
The same is true for the modified contributions to $a_{\omega_i } $ we list
here for arbitrary quark mass.

\vspace{3mm}
\begin{eqnarray*}
a_{\omega_{1}}(d_i\tilde{u}_i \tilde{u}_i) & = & \frac{g^2N_cq_{u_i}}{16\pi^2}
\int_0^1 dx \frac{x(x-1)[b_i(x-2)+x]}{(a_i+\frac{1}{\delta })x+b_i(1-x)-x(1-x)}
\\
a_{\omega_{1}}(d_id_i \tilde{u}_i) & = & \frac{g^2N_cq_{d_i}}{16\pi^2}
\int_0^1 dx \frac{x^2[b_i(x+1)+x-1]}{b_ix+(a_i+\frac{1}{\delta })(1-x)-x(1-x)}
\\
a_{\omega_{1}}(u_i\tilde{d}_i \tilde{d}_i) & = & \frac{g^2N_cq_{d_i}}{16\pi^2}
\int_0^1 dx \frac{x^2(b_i+\frac{1}{\delta })(1-x)}{(b_i+\frac{1}{\delta })x
+a_i(1-x)-x(1-x)} \\
a_{\omega_{1}}(u_iu_i\tilde{d}_i) & = & \frac{g^2N_cq_{u_i}}{16\pi^2}
\int_0^1 dx \frac{x^2(b_i+\frac{1}{\delta })(1-x)}{a_ix+(b_i+\frac{1}{\delta })
(1-x)
-x(1-x)}
\end{eqnarray*}
\vspace{5mm}

As in the exactly supersymmetric case we obtain the corresponding
$a_{\omega_2}$
contributions with the substitutions $q_{d_i} \leftrightarrow -q_{u_i}$,
$a_i \leftrightarrow b_i$.%, with $a_i=(\frac{m_{u_i}}{m_W} )^2$ and
%$b=(\frac{m_{d_i}}{m_W} )^2$.

We have not new contribution to $\Delta K_{WH}$ due to the introduction
of the universal scalar mass because the off diagonal
magnetic transition does not receive any influence
from scalar loops.

\newpage
\begin{center}
\begin{tabular}{|r||r|r|r|r|r||}    \hline
$ \alpha \rightarrow $ & .64 & .44 & .25 & .16 & $10^{-4}$ \\ \cline{1-1}
$\mu \downarrow $    &     &     &     &     &         \\ \hline\hline
$10^{-2}$            & 2.35  & 2.13    & 2.02    & 1.99    & 1.95         \\
$10^{-1}$ & 1.53 & 1.30 & 1.20 & 1.17 & 1.13 \\
.5 & .70 & .47 & .37 & .34 & .30 \\
1 & .33 & .11 & 2$\times 10^{-3}$ & -.03 & -.07 \\
2 & -.02 & -.25 & -.35 & -.38 & -.42 \\
3 & -.21 & -.43 & -.53 & -.57 & -.61 \\
4 & -.33 & -.55 & -.66 & -.69 & -.73 \\
5 & -.42 & -.64 & -.75 & -.78 & -.82 \\
10 & -.66 & -.89 & -.99 & -1.02 & -1.06 \\
20 & -.84 & -1.07 & -1.17 & -1.20 & -1.24 \\
50 & -1.01 & -1.23 & -1.33 & -1.37 & -1.41 \\
100 & -1.08 & -1.31 & -1.41 & -1.44 & -1.48 \\  \hline
\end{tabular}

\vspace{4mm}
\nopagebreak
{\sc Table 1.} $\Delta K_{WW}\times 10^{-3}$ in the Standard Model.
$\mu =(\frac{m_h}{m_W} )^2$,
$\alpha =(\frac{m_W}{m_t} )^2 $.

$m_h$ is the mass of the only physical Higgs scalar in SM.
\end{center}
\vspace{8mm}

\vspace{10mm}
\begin{center}
\begin{tabular}{|r||r|r|r|r|r||}   \hline
$ \alpha \rightarrow $ & .64 & .44 & .25 & .16 & $10^{-4}$ \\  \cline{1-1}
$\mu \downarrow $    &     &     &     &     &         \\ \hline\hline
$10^{-2}$ & 3.06 & 2.89 & 2.83 & 2.81 & 2.81 \\
$10^{-1}$ & 2.73 & 2.56 & 2.50 & 2.49 & 2.48 \\
.5 & 2.46 & 2.28 & 2.22 & 2.21 & 2.21 \\
1 & 2.36 & 2.18 & 2.12 & 2.11 & 2.10 \\
2 & 2.28 & 2.10 & 2.04 & 2.03 & 2.02 \\
3 & 2.24 & 2.06 & 2.00 & 1.99 & 1.98 \\
4 & 2.21 & 2.03 & 1.97 & 1.96 & 1.96 \\
5 & 2.20 & 2.02 & 1.96 & 1.95 & 1.94 \\
10 & 2.16 & 1.98 & 1.92 & 1.91 & 1.90 \\
20 & 2.13 & 1.95 & 1.89 & 1.88 & 1.87 \\
50 & 2.11 & 1.93 & 1.87 & 1.86 & 1.86 \\
100 & 2.11 & 1.93 & 1.87 & 1.86 & 1.85 \\   \hline
\end{tabular}

\vspace{4mm}
\nopagebreak
{\sc Table 2.} $\Delta K_{WW}$ in the minimal exactly supersymmetric version
of the standard model.

Here $m_h$ in
$\mu =(\frac{m_h}{m_W} )^2$ is the mass of the neutral Higgs bosons
which are not in the $Z$ supermultiplet \cite{bil-gas}.
\end{center}

\begin{center}
\begin{tabular}{|r||r|r|r||r|r|r||r|r|r||}    \hline
$\delta \rightarrow$ & & 1 & & & .25 & & & .04 &   \\ \hline
$\mu \downarrow \alpha \rightarrow$ & .64 & .25 & .16 &
.64 & .25 & .16 &
.64 & .25 & .16     \\ \hline\hline
$10^{-2}$ & -2.93 & -3.26 & -3.27 & -3.06 & -3.44 & -3.49 &
-3.44 & -3.79 & -3.83  \\
$10^{-1}$ & -2.59 & -2.91 & -2.93 & -2.68 & -3.06 & -3.11 &
-3.06 & -3.40 & -3.45  \\
.5        & -2.32 & -2.65 & -2.66 & -2.32 & -2.70 & -2.75 &
-2.67 & -3.01 & -3.06  \\
1         & -2.25 & -2.57 & -2.59 & -2.17 & -2.54 & -2.60 &
-2.49 & -2.83 & -2.88  \\
2         & -2.21 & -2.54 & -2.55 & -2.04 & -2.42 & -2.47 &
-2.31 & -2.66 & -2.71  \\
3         & -2.20 & -2.53 & -2.55 & -1.98 & -2.36 & -2.41 &
-2.22 & -2.57 & -2.61  \\
4         & -2.20 & -2.53 & -2.54 & -1.95 & -2.32 & -2.37 &
-2.16 & -2.50 & -2.55  \\
5         & -2.20 & -2.53 & -2.54 & -1.92 & -2.30 & -2.35 &
-2.11 & -2.46 & -2.50  \\
10        & -2.20 & -2.53 & -2.54 & -1.87 & -2.25 & -2.30 &
-1.99 & -2.34 & -2.38  \\
20        & -2.20 & -2.53 & -2.54 & -1.83 & -2.21 & -2.26 &
-1.90 & -2.25 & -2.29  \\
50        & -2.20 & -2.52 & -2.54 & -1.79 & -2.17 & -2.22 &
-1.82 & -2.16 & -2.21  \\
100       & -2.19 & -2.51 & -2.53 & -1.77 & -2.14 & -2.20 &
-1.76 & -2.10 & -2.15  \\ \hline
\end{tabular}

\vspace{10mm}
\nopagebreak
\begin{tabular}{|r||r|r|r||r|r|r||}   \hline
$\delta \rightarrow $ & & .01 & & & .0001 & \\  \hline
$\mu \downarrow \alpha \rightarrow $    & .64 & .25 & .16
 & .64 & .25 & .16  \\ \hline\hline
$10^{-2}$ & -3.61 & -3.94 & -3.97 & -3.72 & -4.04 & -4.08 \\
$10^{-1}$ & -3.22 & -3.56 & -3.59 & -3.33 & -3.66 & -3.69 \\
.5        & -2.83 & -3.16 & -3.20 & -2.94 & -3.27 & -3.30 \\
1         & -2.65 & -2.98 & -3.02 & -2.76 & -3.09 & -3.12 \\
2         & -2.47 & -2.81 & -2.84 & -2.58 & -2.91 & -2.94 \\
3         & -2.38 & -2.71 & -2.74 & -2.49 & -2.81 & -2.85 \\
4         & -2.31 & -2.64 & -2.68 & -2.42 & -2.75 & -2.78 \\
5         & -2.26 & -2.60 & -2.63 & -2.37 & -2.70 & -2.73 \\
10        & -2.13 & -2.46 & -2.50 & -2.24 & -2.57 & -2.60 \\
20        & -2.03 & -2.36 & -2.40 & -2.14 & -2.47 & -2.50 \\
50        & -1.93 & -2.26 & -2.29 & -2.05 & -2.38 & -2.41 \\
100       & -1.87 & -2.20 & -2.23 & -2.00 & -2.33 & -2.36 \\ \hline
\end{tabular}

\vspace{5mm}
\nopagebreak
{\sc Table 3.} $\Delta K_{WW} \times 10^{-3}$ in the MSSM with Susy broken

by an universal scalar mass $\tilde{m}$. $\delta =(\frac{m_W}{\tilde{m}} )^2.$
\end{center}


\newpage

\begin{thebibliography}{99}
\bibitem{fer-por} S. Ferrara and M. Porrati,  Phys. Lett. B288 (1992) 85;
\bibitem{bil-gas} Bilchak, R. Gastmans and A. van Proeyen, Nucl. Phys.
B273 (1986) 46;
\bibitem{wei} S. Weinberg, in Lectures on Elementary Particles and Quantum
Field Theory, Proc. of the Summer Institute, Brandeis University, 1970,
ed. S. Deser (MIT Press, Cambridge, MA, 1970), Vol. I;
\bibitem{fer-tel} S. Ferrara, M. Porrati and V. Telegdi, Phys. Rev. D46 (1992)
3529;
\bibitem{bar-gas} W. A. Bardeen, R. Gastmans and B. Lautrup, Nucl. phys.
B46 (1972) 319;
\bibitem{hag-pec} K. Hagiwara, R. D. Peccei, D. Zeppenfeld and K. Hikasa,
Nucl. Phys. B282 (1987) 253;
\bibitem{rob} R. W. Robinett, Phys. Rev. D31 (1985) 1657;
\bibitem{fer-rem} S. Ferrara and E. Remiddi, Phys. Lett. B53 (1974) 347;
\bibitem{mik-sam} K. O. Mikaelian, M. A. Samuel and D. Brown, Lett. Nuovo
Cim. 27 (1980) 211;
\bibitem{hab-kan} H. E. Haber and G. L. Kane, Phys. Rep. 117 (1985) 75;
\bibitem{lep-col} The LEP Collaborations: ALEPH, DELPHI, L3, OPAL, Phys.
Lett. B276 (1992) 247;
\bibitem{cdf-col} F. Abe et al., The CDF Collaboration, FERMILAB-PUB-94/097-E
submitted to Phys. Rev. D; FERMILAB-PUB-94/116-E submitted to Phys. Rev.
Lett.;
\bibitem{gun-hab} J. F. Gunion and H. E. Haber, Nucl. Phys. B272 (1986) 1;
\bibitem{fer-mas} S. Ferrara and A. Masiero, CERN TH-6846/93, Proc. of
"SUSY 93" Northeastern University, Boston, USA (1993) and Proc of
26th Workshop, Eloisatron Project, "From Superstring to Supergravity",
Erice, 1992;
\bibitem{cou-ng} G. Couture and J. N. Ng, Z. Phys. C - Particles and Fields
35 (1987) 65;
\bibitem{sam-sam} K. B. Samuel, M. A. Samuel and G. Li, Osu Research Note 218,
May, 1989;
\bibitem{iba-lus} L. E. Ib\'{a}\~{n}ez and D. L\"{u}st,
Nucl. Phys. B382 (1992) 305; B. de Carlos, J. A. Casas and C. Mu\~{n}oz,
phys. Lett. B299 (1993) 234;  B. de Carlos, J. A. Casas and C. Mu\~{n}oz,
CERN-TH 6922/93;
\bibitem{sch} J. Schwidling, Proc. of the Int. Europhysics Conference,
Marseille, 1993;
\bibitem{sam-li} D. Zeppenfeld, Phys. Lett. B183 (1987) 380;
U. Baur and D. Zeppenfeld, Nucl. Phys. B325 (1989) 253;
E. N. Argyres et al., Phys. Lett. B272 (1991) 431;
E. N. Argyres et al., Phys. Lett. B280 (1992) 324;
M. A. Samuel, G. Li, N. Sinha, R. Sinha and M. K. Sundaresan,
Phys. Lett. B280 (1992) 124;




\end{thebibliography}
\end{document}